\documentclass[preprint,
superscriptaddress, 
secnumarabic, 
amssymb, 
aps, prb
]{revtex4-2}

\usepackage{bm}
\usepackage{graphicx}
\usepackage{gensymb}
\usepackage{amsmath}
\usepackage{amssymb}
\usepackage{color}
\usepackage{ulem}
\usepackage[driverfallback=dvipdfm]{hyperref}

\hypersetup{%
 setpagesize=false,%
 bookmarks=true,%
 bookmarksdepth=tocdepth,%
 bookmarksnumbered=true,%
 colorlinks=true,%
 allcolors=blue,%
 pdftitle={},%
 pdfsubject={},%
 pdfauthor={},%
 pdfkeywords={}}

\begin{document}

\title{Unusual surface states associated with the $\mathcal{PT}$-symmetry breaking and antiferromagnetic band folding in NdSb}%

\author{Asuka Honma}
%\thanks{These authors contributed equally to this work.}
\affiliation{Department of Physics, Graduate School of Science, Tohoku University, Sendai 980-8578, Japan}

\author{Daichi Takane}
%\thanks{These authors contributed equally to this work.}
\affiliation{Department of Physics, Graduate School of Science, Tohoku University, Sendai 980-8578, Japan}

\author{Seigo Souma}
\thanks{Corresponding authors:\\
s.souma@arpes.phys.tohoku.ac.jp\\
t-sato@arpes.phys.tohoku.ac.jp}
\affiliation{Center for Science and Innovation in Spintronics (CSIS), Tohoku University, Sendai 980-8577, Japan}
\affiliation{Advanced Institute for Materials Research (WPI-AIMR), Tohoku University, Sendai 980-8577, Japan}

\author{Yongjian Wang}
\affiliation{Institute of Physics II, University of Cologne, K$\ddot{o}$ln 50937, Germany}

\author{Kosuke Nakayama}
\affiliation{Department of Physics, Graduate School of Science, Tohoku University, Sendai 980-8578, Japan}
\affiliation{Precursory Research for Embryonic Science and Technology (PRESTO), Japan Science and Technology Agency (JST), Tokyo, 102-0076, Japan}

\author{Miho Kitamura}
\affiliation{Institute of Materials Structure Science, High Energy Accelerator Research Organization (KEK), Tsukuba, Ibaraki 305-0801, Japan}
\affiliation{National Institutes for Quantum Science and Technology (QST), Sendai 980-8579, Japan}

\author{Koji Horiba}
\affiliation{National Institutes for Quantum Science and Technology (QST), Sendai 980-8579, Japan}

\author{Hiroshi Kumigashira}
\affiliation{Institute of Multidisciplinary Research for Advanced Materials (IMRAM), Tohoku University, Sendai 980-8577, Japan}

\author{Takashi Takahashi}
\affiliation{Department of Physics, Graduate School of Science, Tohoku University, Sendai 980-8578, Japan}

\author{Yoichi Ando}
\affiliation{Institute of Physics II, University of Cologne, K$\ddot{o}$ln 50937, Germany}

\author{Takafumi Sato}
\thanks{Corresponding authors:\\
s.souma@arpes.phys.tohoku.ac.jp\\
t-sato@arpes.phys.tohoku.ac.jp}
\affiliation{Department of Physics, Graduate School of Science, Tohoku University, Sendai 980-8578, Japan}
\affiliation{Center for Science and Innovation in Spintronics (CSIS), Tohoku University, Sendai 980-8577, Japan}
\affiliation{Advanced Institute for Materials Research (WPI-AIMR), Tohoku University, Sendai 980-8577, Japan}
\affiliation{International Center for Synchrotron Radiation Innovation Smart (SRIS), Tohoku University, Sendai 980-8577, Japan}
\affiliation{Mathematical Science Center for Co-creative Society (MathCCS), Tohoku University, Sendai 980-8577, Japan}

\date{\today}

\begin{abstract}
We have performed angle-resolved photoemission spectroscopy on NdSb, which exhibits the type-I antiferromagnetism below $T_\mathrm{N} = 16\,\mathrm{K}$. By utilizing microfocused photons, we succeeded in selectively observing the band structure for all three types of single-\textit{q} antiferromagnetic (AF) domains at the surface. We found that two of the three surfaces whose AF-ordering vector lies within the surface plane commonly show twofold symmetric surface states (SSs) around the bulk-band edges, whereas the other surface with an out-of-plane AF-ordering vector displays fourfold symmetric shallow electronlike SS at the Brillouin-zone center. We suggest that these SSs commonly originate from the combination of the $\mathcal{PT}$- (space-inversion and time-reversal) symmetry breaking at the surface and the band folding due to the AF order. The present results pave a pathway toward understanding the relationship between the symmetry and the surface electronic states in antiferromagnets.
\end{abstract}

\maketitle

\section{INTRODUCTION}
The interplay among symmetry, electronic states, and physical properties is a key topic of condensed-matter physics, as highlighted by the formation of charge/spin-density wave (CDW/SDW) triggered by the Fermi surface (FS) nesting accompanied by the change in the translational symmetry, as well as the superconductivity associated with broken gauge symmetry characterized by an energy gap in the low-energy excitation. 
Crystal surface provides a fertile playground to study such interplay, because it inherently breaks the translational and space inversion ($\mathcal{P}$) symmetries, often leading to unconventional physical properties originating from exotic surface states (SSs) such as the giant spin-split Rashba states \cite{LaShellPRL1996, KoroteevPRL2004} and topological Dirac-cone states \cite{HasanRMP2010, QiRMP2011, AndoJPSJ2013}. 
Breaking time-reversal ($\mathcal{T}$) symmetry via ferromagnetism is regarded as a promising route to realize even more exotic SSs, such as massive Dirac-cone states responsible for the quantum anomalous Hall effect \cite{ChangScience2013}. 
Besides the $\mathcal{T}$-symmetry breaking, the variety of magnetic structures and the resulting abundant crystal symmetries, along with their breaking, have been theoretically suggested to enrich the physics in antiferromagnets. 
Characteristic symmetries of antiferromagnets so far discussed are the combination of $\mathcal{T}$ symmetry and the translational symmetry (\textit{S} symmetry), which gives rise to the surface selective massive Dirac cone features predicted in antiferromagnetic (AF) TIs \cite{MongPRB2010} and experimentally studied in the MnBi$_2$Te$_4$ family \cite{OtrokovNature2019, HaoPRX2019, LiPRX2019, ChenPRX2019}. 
Another important symmetry is the combination of $\mathcal{P}$ and $\mathcal{T}$ symmetries ($\mathcal{PT}$ symmetry) that has been recently discussed in magnetic Weyl semimetals \cite{WanPRB2011}, Kagome magnets \cite{LinPRB2020}, and altermagnets \cite{SmejkalSciAdv2020, YuanPRB2020, SmejkalPRX2022}. 
However, in a broader perspective, the interplay between surface electronic states and the combined symmetry in antiferromagnets is still unclear, despite the common interest in utilizing antiferromagnets for realizing emergent quantum phenomena and applications in spintronics \cite{MongPRB2010, OtrokovNature2019, SmejkalSciAdv2020, LiNP2010, NakatsujiNature2015, WadleyScience2016}.

Here, we focus on rare-earth monopnictides RX$_\mathrm{p}$ (R = rare earth; X$_\mathrm{p}$ = pnictogen) with a rocksalt structure [see Fig.\ \hyperref[FIG1]{1(a)}] in which the strong coupling between the electronic states and antiferromagnetism was pointed out \cite{SettaiJPSJ1994, KumiPRB1997, TakayamaJPSJ2009, JangSciAdv2019, OinumaPRB2019, KurodaNC2020, SchrunkNature2022}. 
For example, CeSb was reported to undergo reconstruction of bulk-band structure and FS across $T_\mathrm{N}$ associated with \textit{p}-\textit{f} mixing \cite{SettaiJPSJ1994, KumiPRB1997, TakayamaJPSJ2009, JangSciAdv2019, KurodaNC2020}. 
Also, strong modulation of topological Dirac-cone SS in the AF phase was clarified in CeBi \cite{OinumaPRB2019}. 
Recently, unusual Fermi-arc SS in the AF phase was reported by angle-resolved photoemission spectroscopy (ARPES) of NdBi \cite{SchrunkNature2022}. 
This SS is characterized by the magnetic splitting different from the conventional Rashba/Zeeman splitting, and its origin is discussed in terms of the topology. 
Subsequent studies have clarified the existence of similar SS in other RX$_\mathrm{p}$’s \cite{KushnirenkoPRB2022, MadPRB2022, LiNPJ2023}, consistent with DFT calculations assuming the putative multiple-\textit{q} type AF structures \cite{SchrunkNature2022, LiNPJ2023, WangCP2023}. 
However, the origin of SS and its relation to the topology are still under an intensive debate, mainly due to the presence of multiple AF domains at the surface. 
In particular, it is unclear what type of symmetry in the AF-ordered state is responsible for the emergence of unusual SS.

In this article, we report the ARPES study of NdSb and suggest the existence of SS that originates from the breaking of $\mathcal{PT}$ symmetry and the change in the translational symmetry associated with the AF order. 
This is enabled by the utilization of microfocused ARPES with a small beam spot to distinguish all three types of single-\textit{q} AF domains at the surface. 
We found that, while the surface electronic states at the (001) facet strongly depend on the magnetic ordering vector of AF domains, they are commonly characterized by the band splitting, band folding, and band hybridization. 
We also discuss implications of the present results in relation to the nontrivial topology.

\section{EXPERIMENTS}
Single crystals of NdSb were grown by the flux method using Sn flux. 
Raw materials were mixed in a molar ratio of $\mathrm{Nd}:\mathrm{Sb}:\mathrm{Sn}=1:1:20$ and placed in an alumina crucible. 
The crucible was sealed in an evacuated quartz tube filled with Ar gas of 50 mbar. 
The ampoule was heated to 1100 °C, kept for 10 h, and then cooled to 700 °C  in 160 h. 
The excessive Sn was removed in a centrifuge. Obtained crystals were characterized by x-ray diffraction measurements. 
Soft-x-ray (SX) ARPES measurements were performed with an Omicron-Scienta SES2002 electron analyzer with energy-tunable synchrotron light at BL2 in Photon Factory (PF), KEK. 
We used linearly polarized light (horizontal polarization) of 515–601 eV. 
VUV-ARPES measurements were performed with microfocused vacuum-ultraviolet (VUV) synchrotron light at BL28 in PF \cite{KitamuraRSI2022}. 
We used linearly or circularly polarized light of 55–75 eV. 
The energy resolution for the SX- and VUV-ARPES measurements was set to be 150 and 10–20 meV, respectively. 
Samples were cleaved \textit{in situ} along the (001) plane of the cubic crystal in an ultrahigh vacuum of $1\times10^{-10}$ Torr. 
Prior to the ARPES measurements, the crystal orientation was determined by x-ray Laue backscattering. 
The Fermi level (\textit{E}$_\mathrm{F}$) of samples was referenced to that of a gold film electrically in contact with the sample holder.

\section{RESULTS AND DISCUSSION}
\begin{figure}
\begin{center}
\includegraphics[width=5.5in]{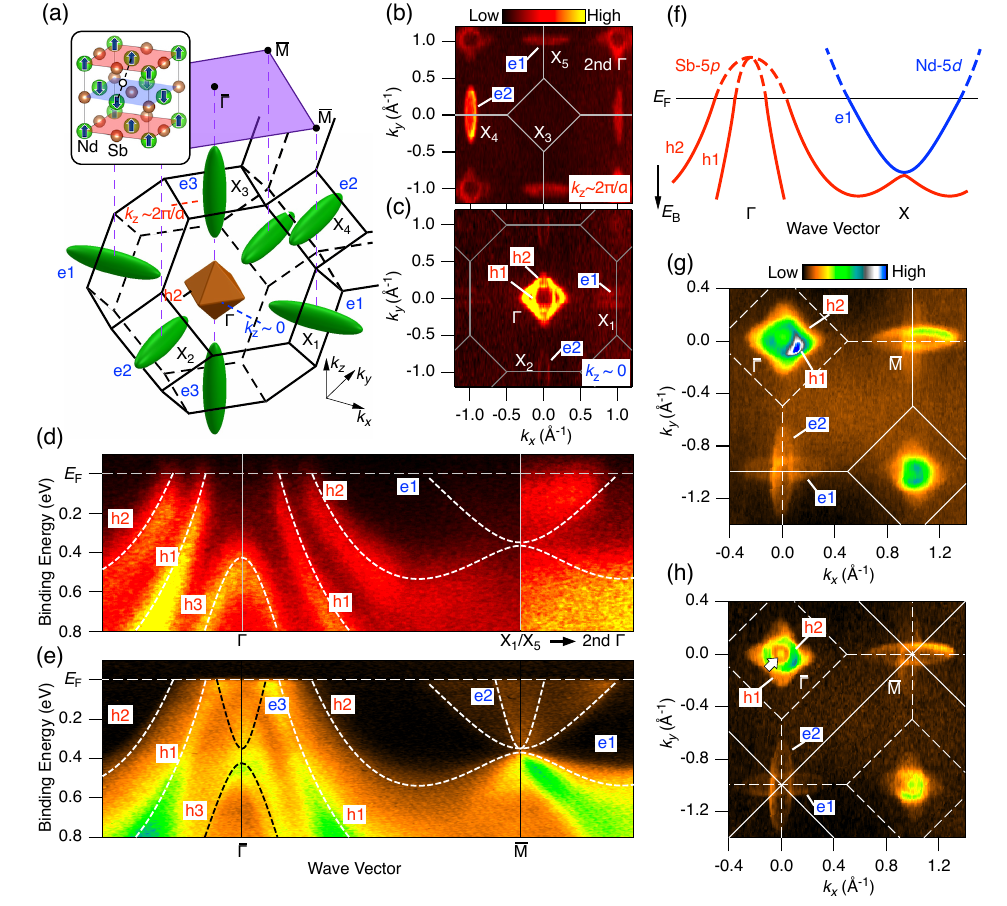}
\hspace{0.2in}
\caption{\label{FIG1}(color online). (a) Schematic FS and bulk fcc BZ of NdSb, together with the surface BZ (purple rectangle). Inset shows the crystal structure of NdSb with the AF structure. $\mathcal{PT}$-preserving center is indicated by open circle. (b), (c) ARPES-intensity mapping at $E_\mathrm{F}$ as a function of in-plane wave vector ($k_x$ and $k_y$) measured at $T = 40\ \mathrm{K}$ at $k_z \sim 2\pi/a$ ($h\nu = 450\ \mathrm{eV}$) and $\sim 0$ ($h\nu = 531\ \mathrm{eV}$), respectively. (d) ARPES intensity along the $\Gamma$X cut plotted as a function of $k_x$ and binding energy measured with SX photons of $h\nu = 531\ \mathrm{eV}$. To enhance the intensity of the e1 band by changing the matrix-element effect, ARPES intensity along another $\Gamma$X cut ($\mathrm{X}_5 - \mathrm{2nd}\ \Gamma$ cut) obtained with $h\nu = 446\ \mathrm{eV}$ is also shown in the right panel. White dashed curves are a guide for the eyes to trace the Sb 5\textit{p} (h1, h2, and h3) and Nd 5\textit{d} (e1) bands. (e) ARPES intensity along the $\bar{\Gamma}\bar{\textrm{M}}$ cut of surface BZ measured at $T = 30\ \mathrm{K}$ with VUV photons of $h\nu = 60\ \mathrm{eV}$. (f) Schematic band diagram of NdSb along the $\Gamma$X cut in the PM phase. (g) ARPES-intensity mapping at $E_\mathrm{F}$ as a function of $k_x$ and $k_y$ at $T = 30\ \mathrm{K}$ measured at $h\nu = 60\ \mathrm{eV}$. (h) Same as (g) but measured at $T = 7\ \mathrm{K}$ in the AF phase. White arrow at the $\bar{\Gamma}$ point indicates SS which appears only in the AF phase.}
\end{center}
\end{figure}

At first, we present the electronic states of NdSb in the paramagnetic (PM) phase. 
By utilizing the bulk-sensitive SX photons to minimize the experimental uncertainty in the out-of-plane wave vector (i.e., \textit{k}$_z$ broadening), we mapped out the FS originating from the bulk bands. 
In-plane FS mapping at $k_z\sim2\pi/a$ plane at $T = 40\ \textrm{K}$ shown in Fig.\ \hyperref[FIG1]{1(b)} signifies an elliptical pocket centered at each X point of the bulk Brillouin zone (BZ), together with a diamondlike pocket at the side $\Gamma$ point (note that the intensity of the pocket at the X$_3$ point appears to be suppressed due to the matrix-element effect). 
The latter diamondlike pocket consists of an outer diamondlike pocket and an inner circlelike one, as better visualized by the FS mapping at the $k_z\sim0$ plane in Fig.\ \hyperref[FIG1]{1(c)}. 
From the ARPES-intensity plot along the $\Gamma$X cut in bulk BZ shown in Fig.\ \hyperref[FIG1]{1(d)}, the pockets centered at the X and $\Gamma$ points are assigned to the Nd 5\textit{d} electron band (e1) and a couple of Sb 5\textit{p} hole bands (h1 and h2), respectively, consistent with the compensated semimetallic nature of the RX$_\mathrm{p}$ family as highlighted by the bulk FS topology in Fig.\ \hyperref[FIG1]{1(a)}.

In RBi, the e1 and h2 bands cross each other midway between $\Gamma$ and X points due to the bulk-band inversion and the hybridization gap due to the strong spin-orbit coupling opens at the intersection. 
On the other hand, in NdSb, the h2 and e1 bands approach at the X point without intersection [see Fig.\ \hyperref[FIG1]{1(d)} and corresponding schematics in Fig.\ \hyperref[FIG1]{1(f)}] and consequently these energy bands keep the noninverted characteristics as in the case of other RSb \cite{OinumaPRB2017}. 
The absence of band inversion is also suggested from the ARPES intensity along the $\bar{\Gamma}\bar{\textrm{M}}$ cut obtained with VUV photons ($h\nu = 60\ \mathrm{eV}$) [see Fig.\ \hyperref[FIG1]{1(e)}], where the band structure obtained with SX photons is overall reproduced, but the band structure at different $k_z$’s ($k_z = 0$ and $2\pi/a$) is simultaneously observed due to the strong $k_z$ broadening caused by the short photoelectron escape depth in VUV measurements \cite{OinumaPRB2017, KumiPRB1998, NayakNC2017}. 
Such $k_z$ broadening is recognized by the observation of the e3 band at the $\bar{\Gamma}$ point and the e2 band at the $\bar{\textrm{M}}$ point both of which originate from the $k_z = 2\pi/a$ plane [see Fig.\ \hyperref[FIG1]{1(a)}], besides the e1, h1, and h2 bands originating from the $k_z = 0$ plane. 

We found no signature of topological Dirac-cone SS at the $\bar{\Gamma}$ and $\bar{\textrm{M}}$ points [see Fig.\ \hyperref[FIG1]{1(e)}] unlike the case of RBi \cite{OinumaPRB2019, NayakNC2017, KurodaPRL2018, LiPRB2018, SakhyaPRB2022}, consistent with the absence of the band inversion, signifying that NdSb is a topologically trivial semimetal in the PM phase. 
The FS mapping in the PM phase shown in Fig.\ \hyperref[FIG1]{1(g)} is consistent with the nontopological semimetallic character because only bulk-derived electron and hole pockets are identified. 
Even when the sample is cooled down to $T = 7\ \mathrm{K}$ well below $T_\mathrm{N}$ ($= 16\ \textrm{K}$), the intensity pattern does not appear to show a significant change, as shown in Fig.\ \hyperref[FIG1]{1(h)}. However, a closer look reveals a qualitative change in the intensity profile; for example, a new tiny pocket appears inside the h1 pocket at the $\bar{\Gamma}$ point (white arrow).

\begin{figure}
\begin{center}
\includegraphics[width=6in]{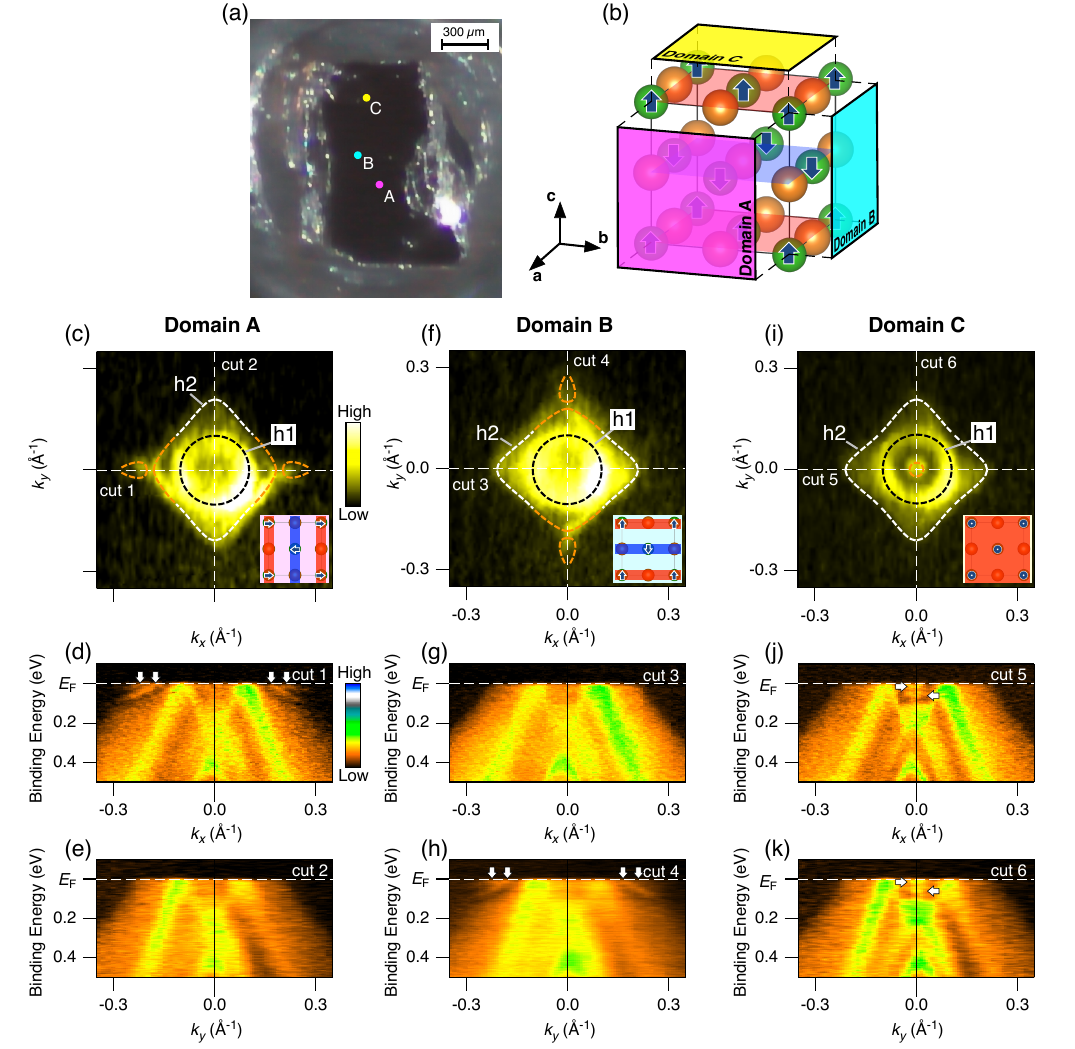}
\hspace{0.2in}
\caption{\label{FIG2}(color online). (a) Optical microscope image of a cleaved surface of NdSb where domain-selective micro-ARPES measurements were performed. (b) Schematic on the relationship between surface orientations and AF structures which signifies three types of AF domains at the surface, called domains A, B, and C. (c) ARPES-intensity mapping at $E_\mathrm{F}$ as a function of $k_x$ and $k_y$ at $T = 8\ \mathrm{K}$ for domain A. (d), (e) ARPES intensity measured along horizontal and vertical $\textbf{k}$ cuts [cuts 1 and 2 in (c)], respectively. (f)--(h) Same as (c)--(e) but for domain B. (i)--(k) Same as (c)--(e) but for domain C.}
\end{center}
\end{figure}

Now we turn our attention to the electronic states in the AF phase. 
As shown in Fig.\ \hyperref[FIG2]{2(a)}, when NdSb forms a single AF domain (e.g. by applying a magnetic field along the [001] axis), the top surface becomes a FM layer with an out-of-plane magnetic moment (called domain C here), whereas two side surfaces have a stripe AF configuration with the magnetic moment parallel to the surface (domains A and B). 
Owing to the cubic symmetry of NdSb, all these domains can be obtained by cleaving the crystal. 
Under the absence of a magnetic field, it is expected that all three types of AF domains as large as a few 100 $\mu$m \cite{KurodaNC2020} coexist on the top surface. 
Thus, by utilizing the microbeam spot of $\sim10\times10\ \mathrm{\mu}\mathrm{m^2}$, one can selectively probe each single AF domain. 
For this sake, we carried out scanning micro-ARPES measurements on the cleaved surface, and were able to resolve all three types of AF domains.

Results of FS mapping around the $\bar{\Gamma}$ point in the AF phase ($T = 8\ \mathrm{K}$) obtained at three representative sample points A--C [Fig.\ \hyperref[FIG2]{2(a)}] are shown in Figs.\ \hyperref[FIG2]{2(c)},\ \hyperref[FIG2]{2(f)}, and\ \hyperref[FIG2]{2(i)}, respectively. 
At point A [Fig.\ \hyperref[FIG2]{2(c)}], one can see the overall $C_2$-symmetric intensity distribution, characterized by the existence of a small pocket on the left- and right-hand sides of the bulk h2 pocket (highlighted by dashed orange circles). 
This pocket can be better recognized from the band dispersion along the horizontal $k_x$ axis (cut 1) in Fig.\ \hyperref[FIG2]{2(d)}, which signifies a couple of shallow bands in the vicinity of $E_\mathrm{F}$ (white arrows). 
These bands are assigned to the SS according to the $h\nu$-dependent ARPES measurements (for details, see Fig.\ \hyperref[FIG6]{6} in Appendix\ \hyperref[apxA]{A}) and the previous ARPES studies of NdBi and NdSb, but the origin is under intensive debate \cite{SchrunkNature2022, KushnirenkoPRB2022, LiNPJ2023}. 
We found that these SS are absent along the vertical $k_y$ axis (cut 2), as seen from both the FS mapping in Fig.\ \hyperref[FIG2]{2(c)} and the band dispersion in Fig.\ \hyperref[FIG2]{2(e)}, suggesting the $C_2$-symmetric nature of the overall electronic structure. 
By changing the measurement geometry, we have confirmed that the observed $C_2$ symmetry is not due to the matrix-element effect; for details, see Fig.\ \hyperref[FIG7]{7} of Appendix\ \hyperref[apxB]{B}.
Since the small pocket is likely associated with the AF-induced band folding as we discuss later, one can attribute the electronic states at the sample point A to the domain A characterized by the AF configuration with the magnetic moment aligned vertically as shown in the inset to Fig.\ \hyperref[FIG2]{2(c)}.

We also identify a counterpart of domain A, namely, domain B with the horizontally aligned AF configuration. 
As shown in Fig.\ \hyperref[FIG2]{2(f)}, the overall FS intensity profile obtained at the sample point B exhibits $C_2$ symmetry but is rotated by 90$^\circ$ with respect to that of domain A [Fig.\ \hyperref[FIG2]{2(c)}]. 
Specifically, small pockets appear along the $k_y$ axis, but not along the $k_x$ axis, as can be identified from the band dispersion along cuts 3 and 4, which signifies the existence of shallow bands only along the $k_y$ cut (cut 4) [Fig.\ \hyperref[FIG2]{2(h)}]. 
We found that the ARPES data at the sample point C, assigned to domain C with the FM top layer showing an out-of-plane magnetic moment, is very different from those of points A and B. 
As shown in Fig.\ \hyperref[FIG2]{2(i)}, the FS mapping shows an overall $C_4$ symmetry as in the case of the PM phase shown in Fig.\ \hyperref[FIG1]{1}, distinct from the $C_2$-symmetric behavior at points A and B in the AF phase [note that the FS mapping in Fig.\ \hyperref[FIG1]{1(h)} was obtained at point C]. 
Small pockets outside the bulk h2 pockets are completely absent, as evident from the plots of band dispersion along cuts 5 and 6 in Figs.\ \hyperref[FIG2]{2(j)} and\ \hyperref[FIG2]{2(k)}. 
All these band/FS features are consistent with the $C_4$-symmetric AF configuration of domain C. 
A careful look at Fig.\ \hyperref[FIG2]{2(i)} reveals a small pocket at the $\bar{\Gamma}$ point which is absent in domains A and B [Figs.\ \hyperref[FIG2]{2(c)} and\ \hyperref[FIG2]{2(f)}]. 
This pocket originates from the complicated reconstruction of band structure at the $\bar{\Gamma}$ point as recognized from the emergence of a couple of shallow features in the vicinity of $E_\mathrm{F}$ around the $\bar{\Gamma}$ point [white arrows in Figs.\ \hyperref[FIG2]{2(j)} and\ \hyperref[FIG2]{2(k)}], distinct from domains A and B. 
These results strongly suggest the successful identification of all three types of AF domains coexisting at the surface. 
The present experimental results definitely rule out the triple-\textit{q} AF structure which requires the existence of a single $C_4$-symmetric AF domain at the (001) surface. 
Moreover, the double-\textit{q} AF structure could also be ruled out because in this case the small pocket should appear in the $C_4$-symmetric manner in domain C \cite{SchrunkNature2022, LiNPJ2023}. 
Therefore, our experimental results strongly suggest the single-\textit{q} nature of the AF structure in NdSb, consistent with the neutron diffraction experiments \cite{NeresonJAP1971, SchobingerJPCSSP1973, ManfrinettiJAC2009}.
It is noted that, to further obtain a direct correspondence between the observed electronic states and AF domains, it is necessary to carry out a polarizing microscopy measurement on a surface at which we performed the domain-selective ARPES measurements; we leave this experiment as a challenge in the future.

To pin down the origin of unusual SS forming the shallow pocket in domains A and B, we have carried out detailed temperature-dependent ARPES measurements across $T_\mathrm{N}$ along cut 1 ($k_x$ axis) for domain A and the result is shown in Fig.\ \hyperref[FIG3]{3(a)}. 
At $T = 8$ K [Fig.\ \hyperref[FIG3]{3(a1)}], one can see a couple of shallow bands crossing $E_\mathrm{F}$ (called S1 and S2 here) outside the highly dispersive bulk h1 band, where the outer S2 band forms a small pocket as seen in Fig.\ \hyperref[FIG2]{2(c)}. 
The S1 and S2 bands significantly reduce their spectral weight on moving away from the $\bar{\Gamma}$ point. 
On increasing temperature, the S1 and S2 bands gradually merge into a single band and become indistinguishable at $T = 13$ K and their spectral weight eventually vanishes at $T \sim 15–16$ K, around $T_\mathrm{N}$ ($= 16$ K). 
This indicates that these bands are associated with the AF transition, as in the case of NdBi \cite{SchrunkNature2022}. 
The AF origin of these bands is also suggested from the observation of discontinuity in the spectral weight across $T_\mathrm{N}$ in the plot of ARPES intensity at $E_\mathrm{F}$ against temperature [Fig.\ \hyperref[FIG3]{3(b)}]. 
To obtain further insights into the overall band structure of S1 and S2, we show in Fig.\ \hyperref[FIG3]{3(c)} the ARPES intensity with enhanced color contrast obtained along the $k_x$ axis in the wider $\mathbf{k}$ range that covers the $\bar{\textrm{M}}$ points at both sides of the $\bar{\Gamma}$ point. 
Obviously, replicas of the S1 and S2 bands also appear around the $\bar{\textrm{M}}$ point (white arrows) and the overall band dispersion of the S1 and S2 bands seems mirror symmetric with respect to $\mathbf{k} = 1/2\bar{\Gamma}\bar{\textrm{M}}$ which is at the AF BZ boundary for domain A. 
It is thus tempting to associate the S1 and S2 bands with the AF-induced band folding. We found that the S1 band around the $\bar{\textrm{M}}$ point smoothly disperses toward higher binding energy up to 0.3 eV at $\mathbf{k} = 1/2\bar{\Gamma}\bar{\textrm{M}}$ and further disperses into the bulk h2 band, although its intensity is suddenly reduced at the intersection with the projection boundary of the bulk e1 pocket. 
This suggests that the intensity of the S1 and S2 bands are enhanced by the surface resonance due to the overlap/proximity with the bulk bands.

Now that the temperature evolution of the SS is established for the $C_2$-symmetric domain A and equivalently for domain B, the next question is whether or not the SS associated with the AF band folding can also be identified for the $C_4$-symmetric domain C. 
To access this issue, we show in Fig.\ \hyperref[FIG4]{4(a)} the temperature dependence of the ARPES intensity obtained along cut 5 ($k_x$ axis) for domain C [Fig.\ \hyperref[FIG2]{2(c)}]. 
At $T = 18$ K in the PM phase [Fig.\ \hyperref[FIG4]{4(a6)}], one can see a rather simple band structure, i.e., a bulk hole band h3 topped at 0.4 eV, a bulk electron band e3 bottomed at $\sim$0.3 eV, besides weaker hole bands h1 and h2. 
At $T = 14$ K [Fig.\ \hyperref[FIG4]{4(a4)}], a new hole band emerges (blue arrow). 
On decreasing temperature, this band systematically moves upward, whereas the bulk h3 band is stationary against the temperature variation. 
This band is assigned to the SS and separates from the bulk bands h3, as seen from the temperature dependence of EDCs in Fig.\ \hyperref[FIG4]{4(b)}. 
A high-resolution ARPES image at a lower temperature ($T = 6.5$ K) shown in Fig.\ \hyperref[FIG4]{4(c)} signifies that this SS consists of a couple of hole bands called S5 and S6, as also evident from the numerical fitting to the momentum distribution curve (MDC) at $E_\mathrm{B} = 0.32$ eV, which crosses the S5 and S6 bands [see Fig.\ \hyperref[FIG4]{4(d)}]. 
Another important aspect of the band structure in the AF phase is the emergence of shallow electron bands as seen in Figs.\ \hyperref[FIG2]{2(j)} and\ \hyperref[FIG2]{2(k)}. 
These bands are assigned to other SSs called S3 and S4 and they survive at least up to 12 K [Figs.\ \hyperref[FIG4]{4(a1)--4(a3)}]. 
On increasing temperature, these bands gradually smear out and eventually vanish at $T = 16$ K, confirming their AF origin. 
From a comparison of extracted band dispersion between the AF phase ($T = 6.5$ K) and PM phase ($T = 18$ K), we concluded that totally four SSs (S3--S6) appear in the AF phase in domain C. 
It is noted that all these SSs cannot be explained in terms of the bulk bands associated with the AF band folding, because there exists no counterpart of bulk bands at the X point [X$_3$ in Fig.\ \hyperref[FIG1]{1(a)}] in the corresponding energy region [see also Fig.\ \hyperref[FIG1]{1(f)}].

Based on the established band structure in the AF phase of all three domains, we discuss the comprehensive feature of SS in more detail. 
Since all the S1--S6 bands are observed only below $T_\mathrm{N}$, they are definitely associated with the AF order. 
Here we propose a phenomenological framework that considers three key band modulations in the AF phase, i.e., (i) the split-off SS from the bulk bands, (ii) the band folding of SS due to the AF potential, and (iii) the hybridization between the original and backfolded SS. 
Although these band modulations occur simultaneously and are essentially indistinguishable, it would be useful to introduce a simplified picture that separates out these effects so as to intuitively understand our observation. 
As shown in Fig.\ \hyperref[FIG5]{5(a)}, the original band structure along the $\bar{\Gamma}\bar{\textrm{M}}$ cut in the PM phase is characterized by the bulk hole bands (h1--h3) at the $\bar{\Gamma}$ point and the bulk electron band (e1) at the $\bar{\textrm{M}}$ point. 
The AF order triggers the emergence of SS detached from the original bulk bands. 
The hole SS with the Sb 5\textit{p} character further shows a spin splitting [black curves in Fig.\ \hyperref[FIG5]{5(b)}], which is suggested from our observation that the SS always appear side by side (S1 and S2, S3 and S4, and S5 and S6). 
We discuss the mechanism of the spin splitting later. 
Since the SSs also suffer from the influence of the AF potential, band folding and resultant band hybridization would take place [Figs.\ \hyperref[FIG5]{5(c)}--\hyperref[FIG5]{5(f)}]. 
For domain A and equivalently domain B, since the folding of SS occurs within the surface plane with respect to the AF BZ boundary of 1/2$\bar{\Gamma}\bar{\textrm{M}}$ [Fig.\ \hyperref[FIG5]{5(c)}], the overall SS band dispersion becomes mirrorsymmetric with respect to $\mathbf{k} = 1/2\bar{\Gamma}\bar{\textrm{M}}$ as demonstrated in Fig.\ \hyperref[FIG3]{3(i)}. 
In addition, the hybridization between hole and electron SS opens an energy gap at each intersection, resulting in the formation of S1 and S2 bands [Fig.\ \hyperref[FIG5]{5(d)}]. 
Although many of the SSs are experimentally obscured owing to their weak intensity [highlighted by gray curves in Fig.\ \hyperref[FIG5]{5(d)}], some surface bands inside the projection of bulk bands are clearly detected in the present experiment because of the intensity enhancement by the surface resonance. 
We found that this picture can also be applied to domain C. 
As shown in Figs.\ \hyperref[FIG5]{5(e)} and\ \hyperref[FIG5]{5(f)}, the SS at the $\bar{\Gamma}$ point which are split off from the bulk e3 and h3 bands hybridize with each other due to the AF-induced out-of-plane band folding to produce the complicated SS dispersion (S3--S6), although the reproduction of all the fine structures needs further elaboration utilizing first-principles band calculations. 

What is the physical mechanism behind the present observation? 
To answer this question, the following two key issues must be clarified: (i) the origin of spin splitting in the SS and (ii) the reason why the SS appear only in the AF phase. 
First, we discuss the issue (i). 
In ordinary antiferromagnets with zero internal magnetic field, the bulk bands are spin degenerate because the $\mathcal{PT}$ symmetry corresponding to the operation to reverse the spin while keeping the same momentum \textbf{k} [$E(\mathbf{k},\uparrow) = E(\mathbf{k},\downarrow)$] holds in the bulk \cite{YuanPRB2020}. 
Since the single-\textit{q} AF phase of NdSb has a $\mathcal{PT}$-preserving center in the bulk crystal [black circle in Fig. \ \hyperref[FIG1]{1(a)}], the $\mathcal{PT}$ symmetry holds and the bulk bands are spin degenerate in both the PM and AF phases. 
On the other hand, at the surface of NdSb in the AF phase, the $\mathcal{PT}$ symmetry is broken at the surface boundary and thereby the energy bands are spin split. 
This splitting is likely associated with the \textit{k}-dependent magnetic exchange interactions because the splitting sets in at $T_\mathrm{N}$ [see Figs.\ \hyperref[FIG3]{3(b)} and\ \hyperref[FIG4]{4(b)}]. 
In this respect, it is interesting to note the similarity to altermagnets where the band splitting due to the $\mathcal{PT}$-symmetry breaking was recently proposed \cite{SmejkalPRX2022}.

Next, we discuss the issue (ii) of why the SSs appear only in the AF phase. 
Obviously, the SSs are not a conventional SS associated with dangling bonds and/or surface relaxation. 
The possibility of surface reconstruction is unlikely because no such reconstruction was observed. 
The possibility of the topological Fermi-arc SS associated with the magnetic Weyl semimetal phase \cite{WanPRB2011} is also excluded because the bulk bands are spin degenerate. 
The $Z_2$ topology associated with the combined \textit{S} symmetry (time reversal and fractional translation) in the AF-TI phase \cite{MongPRB2010} is further excluded because NdSb has no band inversion and the Dirac-cone SS is absent (Fig.\ \hyperref[FIG1]{1}). 
It is noted that, since the magnetic space group ($\#$128.140) of the single-\textit{q} AF structure is a subgroup of that of the PM phase ($\#$225) \cite{PerezARMR2015, MSGindentifer}, there is no new symmetry operation activated only in the AF phase. %
%\Erase{Hence, the emergence of the SS implies the change in the hidden topology in the bulk elec}
%\Erase{tronic states. 
%This change may be triggered by the band folding and hybridization, as seen}
%\Erase{ from the behavior of SS discussed in Fig.\ \hyperref[FIG5]{5}.}
%\Erase{We thus suggest}
We suggest that the essential ingredients to realize the SS of NdSb in terms of symmetry are the combination of (i) the TRS breaking, (ii) the $\mathcal{PT}$-symmetry breaking, and (iii) the change in the translational symmetry.

We emphasize that the surface of antiferromagnets naturally satisfies all the above three conditions, whereas the other materials in the ordered phase such as ferromagnets, ferroelectrics, and CDW compounds only partially satisfy these conditions. 
It is highly desirable to search for exotic SS associated with the entanglement of multiple symmetries in other antiferromagnets. 
Such SS would show a Berry phase effect seen in ferromagnets and recently also in an altermagnet due to the spin splitting of energy bands \cite{SmejkalPRX2022, XiaoRMP2010} and has a potential to lead to the exotic surface properties such as surface anomalous Hall conductivity, surface anomalous Nernst effect, and surface magneto-optical response \cite{SmejkalSciAdv2020, ChenPRL2014, NakatsujiNP2017, SamantaJAP2020}. 
It is thus desirable to investigate the relationship between the spectroscopically identified exotic SS and the surface anomalous physical properties in antiferromagnets.

\section{CONCLUSION}
The AF-domain-selective micro-ARPES measurements of NdSb have clarified the existence of three types of surface electronic states characterized by the $C_2$ and $C_4$ symmetries corresponding to the single-\textit{q} AF domains with in-plane and out-of-plane magnetic moments, respectively. 
We found that both the $C_2$- and $C_4$-symmetric surfaces are characterized by the emergence of new SS in the AF state. 
Although the SS appears in different locations of the surface BZ, their origins are commonly explained in terms of the breaking of $\mathcal{PT}$ symmetry and the AF band folding. 
The present result lays a foundation for studying exotic surface properties of antiferromagnets.

\textit{Note added:}  Recently, we became aware of a similar work by Y. Kushnirenko \textit{et al.} \cite{KushnirenkoArXiv2023}, which reports the domain-selective micro-ARPES study of NdSb.

\begin{acknowledgments}
We thank Y. Kubota, T. Kato, T. Kawakami, and N. Watanabe for their assistance in the ARPES experiments. 
This work was supported by JST-CREST (No. JPMJCR18T1), JST-PRESTO (No. JPMJPR18L7), and Grant-in-Aid for Scientific Research (JSPS KAKENHI Grant No. JP21H04435 and No. JP19H01845), Grant-in-Aid for JSPS Research Fellow (No. JP23KJ0210 and No. JP18J20058), KEK-PF (Proposal No. 2021S2-001 and No. 2022G652), and UVSOR (Proposal No.21-658 and No. 21-847). 
The work in Cologne was funded by the Deutsche Forschungsgemeinschaft (DFG, German Research Foundation) - Project number 277146847 - CRC 1238 (Subproject A04). A.H. thanks GP-Spin and JSPS, and D.T. thanks JSPS and Tohoku University Division for Interdisciplinary Advanced Research and Education.
\end{acknowledgments}

\clearpage

\appendix
\section{PHOTON-ENERGY DEPENDENCE}
\label{apxA}To determine the $k_z$ dispersion and clarify the bulk/surface nature of energy bands, we have carried out $h\nu$-dependent ARPES experiments.
We show in Figs.\ \hyperref[FIG6]{6(a)} and\ \hyperref[FIG6]{6(b)} some representative ARPES data at different $h\nu$'s (55–75 eV) in the $k_z$ range covering the $\Gamma\textrm{X}$ length of bulk BZ for domains A and C, respectively.
We have confirmed that the energy position of the S1 and S2 bands (defined in Fig.\ \hyperref[FIG3]{3}) for domain A and the S3--S6 bands (defined in Fig.\ \hyperref[FIG4]{4}) for domain C are unchanged even upon $h\nu$ variation, showing their surface origin (see dashed curves), in contrast to bulk bands, h1 and h2.

\section{MATRIX ELEMENT EFFECTS}
\label{apxB}
\begin{figure}
\begin{center}
\includegraphics[width=6in]{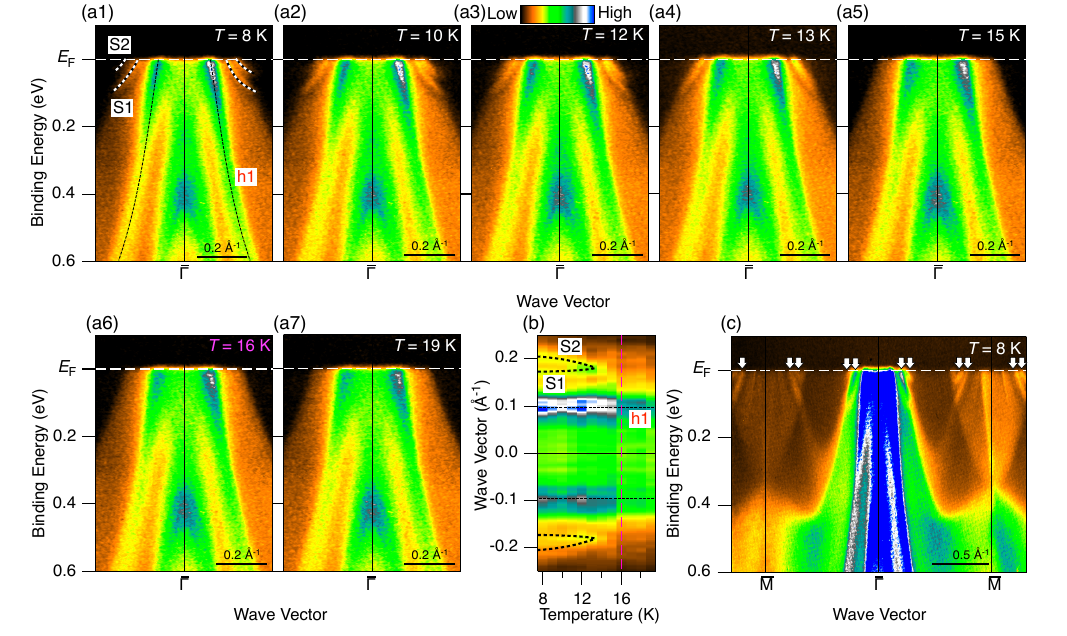}
\hspace{0.2in}
\caption{\label{FIG3}(color online). (a) Temperature dependence of ARPES intensity in NdSb along the $k_x$ cut [cut 1 in Fig. 2(c)] for domain A measured with $h\nu = 55\ \mathrm{eV}$. (b) ARPES intensity at $E_\mathrm{F}$ along the $k_x$ cut plotted against temperature. Dashed curves are a guide for the eyes to highlight S1, S2, and h1. (c) ARPES intensity in the AF phase along the $\bar{\textrm{M}}\bar{\Gamma}\bar{\textrm{M}}$ cut for domain A with enhanced color contrast. White arrows indicate the $k_\mathrm{F}$ points of SS.}
\end{center}
\end{figure}

\begin{figure}
\begin{center}
\includegraphics[width=6in]{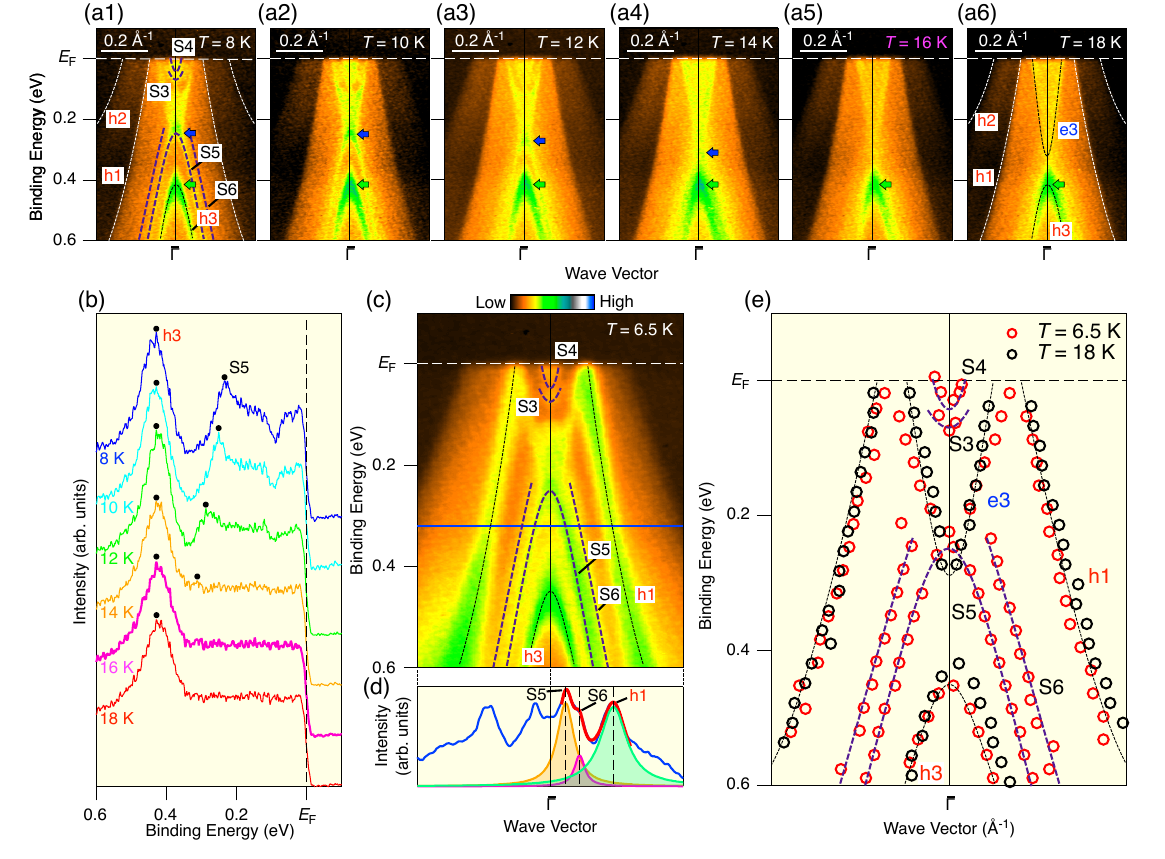}
\hspace{0.2in}
\caption{\label{FIG4}(color online). (a) Temperature dependence of ARPES intensity in NdSb measured along the $k_x$ cut [cut 5 in Fig. 2(i)] for domain C measured with $h\nu = 65\ \mathrm{eV}$. Green and blue arrows represent the top of h3 and split-off S5 bands, respectively. (b) Temperature dependence of EDCs at the $\bar{\Gamma}$ point. Black dots indicate the peak position of S5 and h3 bands. (c) High-resolution ARPES intensity obtained around the $\bar{\Gamma}$ point at $T = 6.5\ \mathrm{K}$ measured with $h\nu = 65\ \mathrm{eV}$. (d) MDC at $E_\mathrm{B} = 0.32$ eV indicated by the blue line in (c). Result of numerical fittings with multiple Lorentzian peaks is also shown in the MDC plot. (e) Experimental band dispersion (dashed curves) extracted from the energy position of bands (open circles) estimated by numerical fittings of MDCs in the AF phase ($T = 6.5\ \mathrm{K}$; red and purple circles/curves) and the PM phase ($T = 18\ \mathrm{K}$; black circles/curves).}
\end{center}
\end{figure}

\begin{figure}
\begin{center}
\includegraphics[width=4.2in]{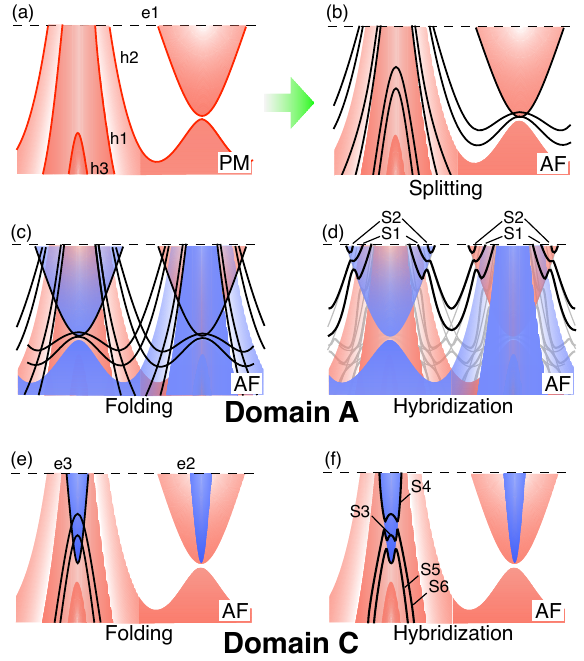}
\hspace{0.2in}
\caption{\label{FIG5}(color online). (a) Schematics of bulk-band structure (red shade) in the PM phase. (b) Schematics of SS (black curves) split off from the bulk h1–h3 bands. (c) Schematics of surface and bulk band structure for domain A that takes into account the AF-induced band folding of bulk and SS. (d) Same as (c) but takes into account band hybridization besides the band folding. SS in which the corresponding feature is resolved by ARPES is highlighted by black curves, whereas other SS which were not clearly observed by ARPES are indicated by gray curves. (e), (f) Same as (c) and (d) but for domain C.}
\end{center}
\end{figure}

\begin{figure}
\begin{center}
\includegraphics[width=5.5in]{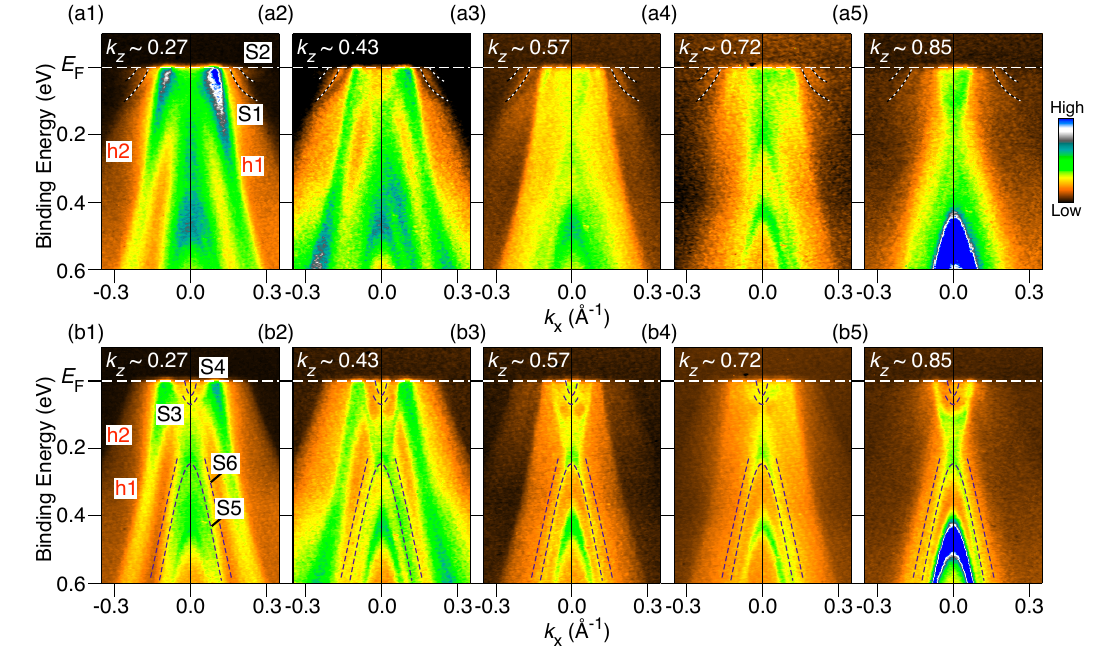}
\hspace{0.2in}
\caption{\label{FIG6}(color online). (a), (b) $h\nu$ dependence ($h\nu = 55–75\ \mathrm{eV}$, every 5 eV step) of the ARPES intensity along the $k_x$ cut for domains A and C, respectively, measured at $T = 7\ \mathrm{K}$. The $k_z$ value in the unit of $2\pi/a$ for the fcc BZ is indicated. Dashed curves highlight the energy dispersion of the surface bands S1--S6.}
\end{center}
\end{figure}

\begin{figure}
\begin{center}
\includegraphics[width=5.5in]{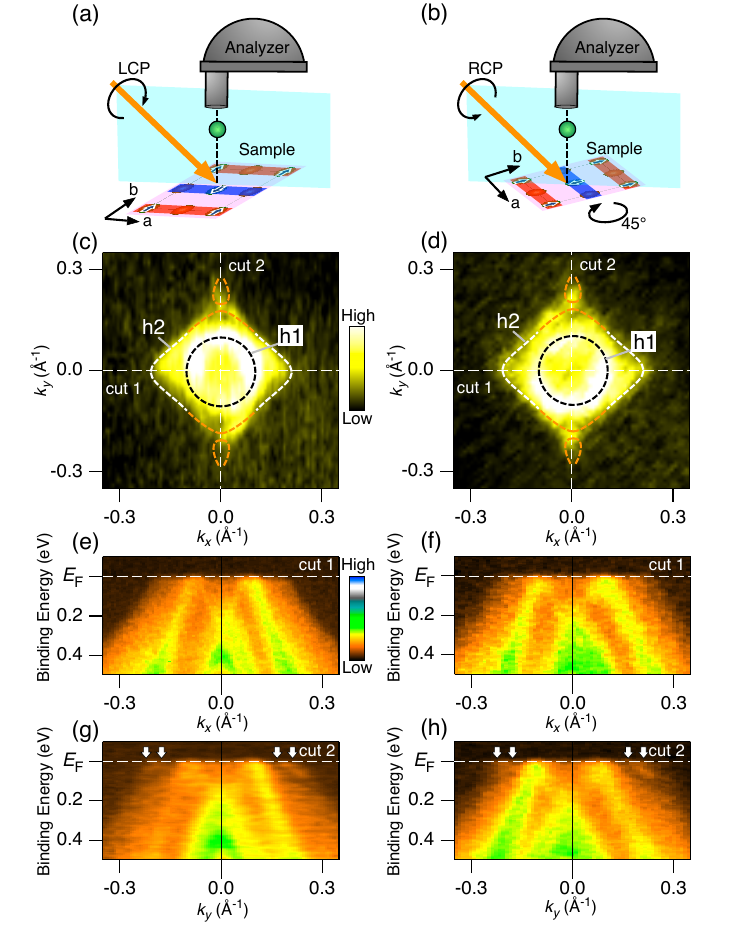}
\hspace{0.2in}
\caption{\label{FIG7}(color online). (a), (b) Schematics of two measurement geometries with different sample azimuth angles and light polarizations (left-handed circular polarization, LCP, and right-handed circular polarization, RCP). (c), (d) FS mapping as a function of $k_x$ and $k_y$ at the experimental geometry shown in (a) and (b), respectively, measured at $T = 8\ \mathrm{K}$ with $h\nu = 60\ \mathrm{eV}$ for domain B. (e), (f) ARPES intensities along cut 1 ($k_x$ axis) shown in (c) and (d), respectively. (g), (h) Same as (e) and (f), but along cut 2.}
\end{center}
\end{figure}

We show in Fig.\ \hyperref[FIG7]{7} representative ARPES data for domain B in the AF phase ($T = 8\ \mathrm{K}$) measured with two different sample geometries and light polarizations at fixed photon energy ($h\nu = 60\ \mathrm{eV}$) as schematically shown in Figs.\ \hyperref[FIG7]{7(a)} and\ \hyperref[FIG7]{7(b)}. Comparing the FS mapping in Figs.\ \hyperref[FIG7]{7(c)} and\ \hyperref[FIG7]{7(d)}, one can commonly recognize small pockets (S2; defined in Fig.\ \hyperref[FIG3]{3}) outside the bulk h2 pocket along the vertical axis ($k_y$ axis), whereas such pockets are absent along the horizontal axis ($k_x$ axis). 
This supports the $C_2$-symmetric electronic states irrespective of the sample geometry and light polarization. 
Such a $C_2$ symmetric nature is also confirmed by directly looking at the corresponding band dispersion along the vertical and horizontal \textbf{k} cuts (cuts 1 and 2, respectively) in Figs.\ \hyperref[FIG7]{7(e)}--\hyperref[FIG7]{7(h)}, where the S1 and S2 bands show up only along the vertical \textbf{k} cut. These results suggest that the $C_2$-symmetric band structure is the intrinsic nature of NdSb in the AF phase and is not due to the matrix-element effect.

\bibliographystyle{prsty}

\clearpage

\clearpage

\clearpage

\clearpage

\clearpage

\clearpage

\clearpage

\end{document}